# Magnetic properties of bismuth-cobalt oxides doped by erbium

N.I. Matskevich, E.V. Korotaev

Nikolaev Institute of Inorganic Chemistry SB RAS, Novosibirsk, Russia

**Abstract**

We synthesized bismuth - cobalt oxide doped by erbium with general formula $Bi_{3-x}Er_xCoO_{3-y}$. Compound has structure of delta-form bismuth oxide. Magnetic properties of the compound were measured by Faraday's method using quartz scales in the temperature range of 80-500 K. The magnetic susceptibility and effective magnetic moment were calculated.

**Introduction**

Compounds on the basis of bismuth oxide and rare-earth elements are multifunctional materials. They are perspective for using as ecological pure inorganic pigments, ceramic oxide generators, etc. [1-5]. One of the important application is based on high ionic conductivity of bismuth oxide delta-form. Compounds with the $\delta$-$Bi_2O_3$ structure stabilized by a variety of elements can exhibit very high oxide ion conductivity and electrocatalytic activity at temperatures below 800 K [6–10]. Moreover, these phases have lower synthesis temperatures than stabilized zirconia phases.

To synthesize compounds with delta-form of bismuth oxide the pure $Bi_2O_3$ is substituted by isovalent and not isovalent elements. Rare-earth metals are doped as isovalent elements, and tungsten, vanadium, calcium, zinc and other elements are used as not isovalent elements.

In present paper the compounds on the basis of bismuth, cobalt and erbium oxides were prepared. Magnetic properties of the compounds were measured in the temperature range of 80-500 K.

**Experimental and results**

Compounds with general formula $Bi_{3-x}Er_xCoO_{3-y}$ were synthesized from bismuth oxide ($Bi_2O_3$), cobalt oxide ($Co_3O_4$) and erbium oxide ($Er_2O_3$) by solid state reactions. Before synthesis bismuth and erbium oxides were treated up to constant weight at 900 K. Then oxides were mixed and repeatedly frayed. After that the mixture was heated at temperature more than 900 K.

Procedures of mixing and heating were repeated many times. Compounds were identified by X-ray phase and X-ray structural analyses (Siemens D5000, Cu K_1, primary beam Ge monochromator). X-ray diffraction pattern is presented in Fig.1. Compound has cubic structure (delta-form of bismuth oxide). Using program FullProf the lattice parameter was determined: a = 5.5182 A.

Measurements of magnetic properties were performed using Faraday method based on quartz scales. The sensitivity of quartz scales is ~0.3 mkg. The temperature range is 80-500 K.

Temperature stabilization during measurements with accuracy ~ 0.15 K is carried out by means of izodrom temperature regulator of PIT-3B. The block of stabilization of magnetic field allows one to perform measurements at magnetic field values from 0 up to 10 k oersted (accuracy of stabilization is 2%). During the measurements the sample was placed in inert atmosphere helium at pressure 5 mmHg.

Static magneto chemistry is one of the main methods to investigate magnetic properties of chemical compounds. Performing temperature investigations of samples magnetic susceptibility $\chi = f(t)$ allows one to reveal point of magnetic phase transactions (ferromagnetic-paramagnetic, anti-ferromagnetic-paramagnetic, etc.), to determine type of magnetic streamlining of paramagnetic centers (ferromagnetic, anti-ferromagnetic), to estimate values of effective magnetic moments of atoms and their spin state. Investigations of magnetic fields dependences $\chi = f(H)$ allows one to reveal existence of distinctive ferromagnetic impurities, to estimate sizes of saturation magnetization for ferromagnetic samples.

For our measurements the intensity of a magnetic field at getting temperature dependences $\chi$ was 9 k oersted.

Obtained experimental dependences of $1/\chi$ in high temperature field is linear from temperature that corresponds to Curie-Veis low in form (Fig. 2):

$\chi = C/(T-\theta)$, where C – Curie constant, θ- Veis constant. Obtained values allows one to calculate effective magnetic moment as $\mu_{eff} = (8 \cdot C)^{1/2}$ which corresponding to $\lim_{T\to\infty}( \mu_{э\ eff} = (8 \cdot \chi \cdot T)^{1/2})$.

We obtained for our compound: C = 3.22 (3) K·sm$^3$/mol, θ = -62 (3) K, $\mu_{eff}$ = 5.07(2) M.B.

**Conclusions**

We synthesized bismuth - cobalt oxide doped by erbium with general formula Bi$_{3-x}$Er$_x$CoO$_{3-y}$. Compound has structure of delta-form bismuth oxide. Magnetic properties of the compound were measured by Faraday's method using quartz scales in the temperature range of 80-500 K. The magnetic susceptibility and effective magnetic moment were calculated.

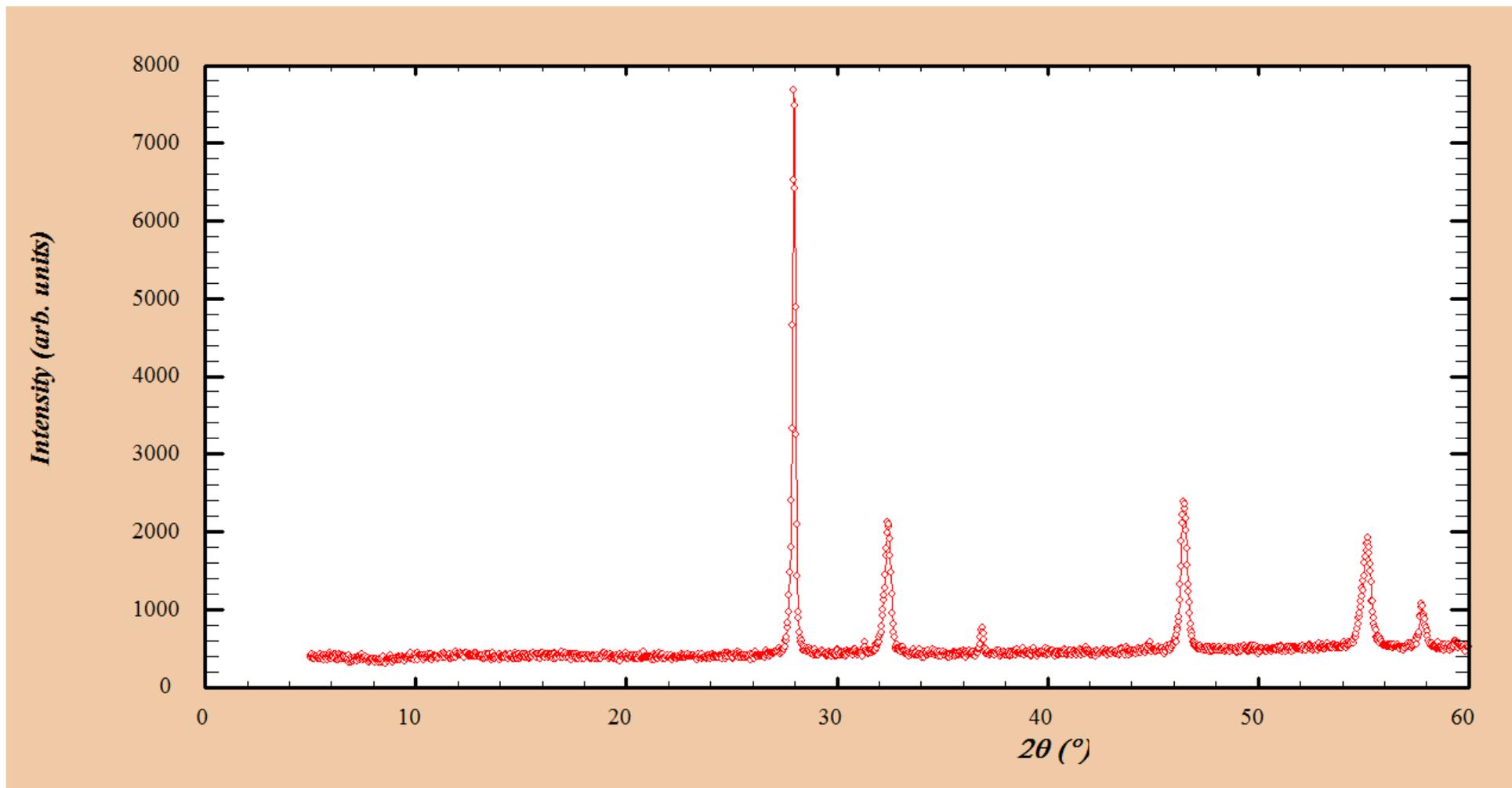

Fig. 1. X-ray diffraction pattern for $Bi_{3-x}Er_xCoO_{6-y}$

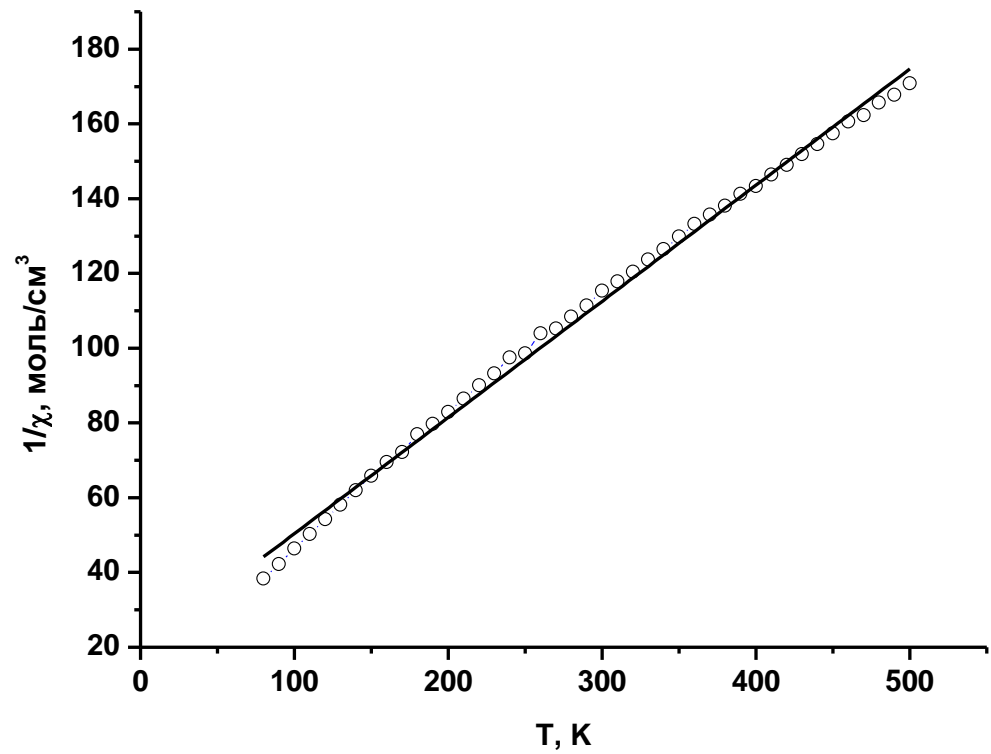

Fig. 2. Dependence of 1/χ (mol/sm³) from temperature